\documentclass[%
 reprint,
 amsmath,amssymb,
 aps,
]{revtex4-2}

\usepackage{graphicx}
\usepackage{dcolumn}
\usepackage{bm}
\usepackage{xcolor}
\usepackage{pdfpages}
\usepackage{pgffor}  

\makeatletter
\AtBeginDocument{\let\LS@rot\@undefined}
\makeatother

\def\supplementfilename{Supplement.pdf}

\pdfximage{\supplementfilename}
\def\numbersupplementpages{\the\pdflastximagepages}

\newif\ifarXiv
\arXivtrue 

\begin{document}

\title{Spin injection and detection in all-van der Waals 2D devices}

\author{Jan Bärenfänger$^1$}
\author{Klaus Zollner$^2$}
\author{Lukas Cvitkovich$^2$}
\author{Kenji Watanabe$^4$}
\author{Takashi Taniguchi$^4$}
\author{Stefan Hartl$^1$}
\author{Jaroslav Fabian $^2$}
\author{Jonathan Eroms$^1$}
\author{Dieter Weiss$^1$}
\author{Mariusz Ciorga$^{1, 3}$}

\affiliation{%
1. Institute for Experimental and Applied Physics, University of Regensburg, Germany
}%

\affiliation{%
2. Institute for Theoretical Physics, University of Regensburg, Germany
}%

\affiliation{
3. Department of Experimental Physics, Faculty of Fundamentals Problems of Technology, Wrocław University of Science and Technology, Poland
}%
\affiliation{
4. National Institute for Materials Science, Tsukuba, Japan
}%

\date{\today}

\begin{abstract}
 In this work we report efficient out-of-plane spin injection and detection in an all-van der Waals based heterostructure using only exfoliated 2D materials. We demonstrate spin injection by measuring spin-valve and Hanle signals in non-local transport in a stack of Fe$_3$GeTe$_2$ (FGT), hexagonal boron nitride (hBN) and graphene layers. FGT flakes form the spin aligning electrodes necessary to inject and detect spins in the graphene channel. The hBN tunnel barrier provides a high-quality interface between the ferromagnetic electrodes and graphene, eliminating the conductivity mismatch problem, thus ensuring efficient spin injection and detection with spin injection efficiencies of up to $P=40$\,\%. Our results demonstrate that FGT/hBN/graphene heterostructures form a promising platform for realizing 2D van der Waals spintronic devices.

\end{abstract}

\maketitle


\section{Introduction}
Combining two-dimensional (2D) materials into van der Waals (vdW) heterostructures opens up new possibilities to study interesting physical phenomena and to develop new device concepts \cite{Novoselov.2016}. Adding magnets to the rich library of 2D materials, comprising metals, insulators, semiconductors and topological insulators, has invigorated the field of spintronics.\cite{Sierra2021,Li2022} One of the key issues in spintronics is the generation of spin polarization in non-magnetic materials.\cite{Fabian2007} A very efficient way of generating spin polarization is electrical spin injection from ferromagnetic materials. The first reports on electrical spin injection and detection in graphene were published in 2007 by Tombros~\textit{et al.} utilizing conventional ferromagnetic Co electrodes with an in-plane magnetization direction and oxide tunnel barriers, and demonstrated the potential of graphene as a spin transport medium with spin relaxation lengths of up to 2\,\textmu m \cite{Tombros.2007}. Since then the foundations of spin transport in graphene have been established, including the role of proximity effects \cite{Zollner.2016, Zollner.2022}, in order to enhance and manipulate the spin signal \cite{Sierra2021}. The discovery of metallic 2D ferromagnets \cite{Deng2018} enables the creation of spintronic devices made entirely of vdW materials. Recently, a report on spin injection from Fe$_{5}$GeTe$_{2}$ into graphene was published, where the spin signal was detected using an electrode made of the conventional ferromagnet Co \cite{Zhao.2023}. All-vdW spin injection devices have also been demonstrated, but with very low efficiency, due to the lack of a tunnel barrier between Fe$_{3}$GeTe$_{2}$ and graphene \cite{He.2023}. 

In this paper, we report on high-efficiency spin injection in all-vdW spin injection devices with a hexagonal boron nitride (hBN) tunnel barrier between Fe$_{3}$GeTe$_{2}$ (FGT) and monolayer graphene. We observe clear spin signals in the spin valve and Hanle measurements, from which we determine the spin injection efficiency, spin relaxation times, and spin diffusion constants. In order to evaluate the experimental results, DFT calculations were conducted.

\section{Experimental details}

\begin{figure*}
\includegraphics[keepaspectratio, width=\textwidth]{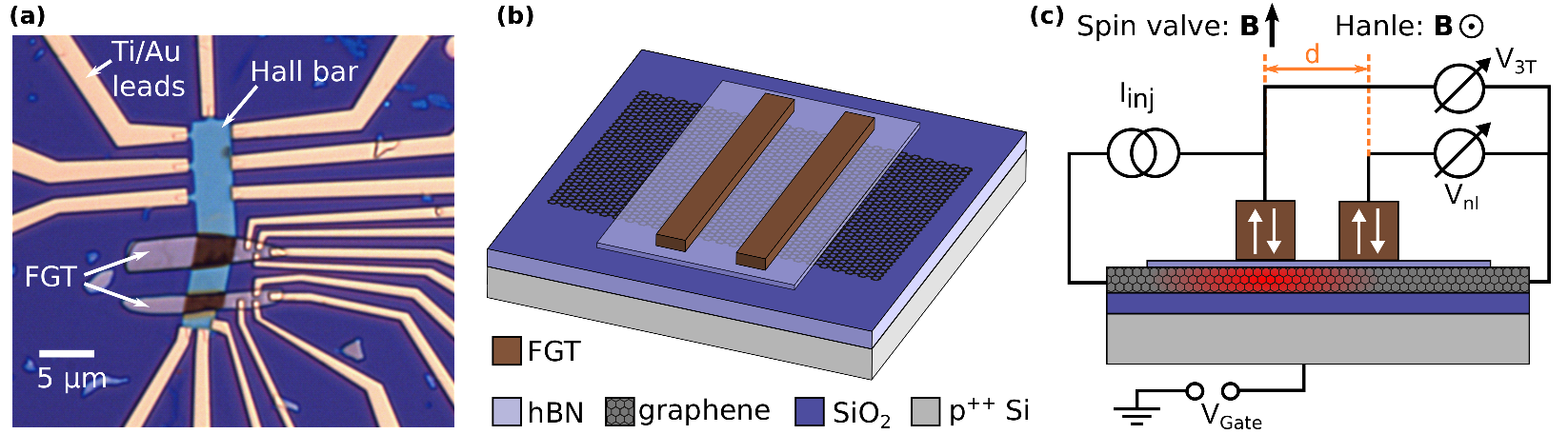}
\caption{(a) Optical micrograph of sample A. (b) Schematic of the samples. The encapsulating hBN is omitted for clarity. (c) Schematic of the non-local measurement setup. For the spin valve (Hanle) measurements the external magnetic field was swept out-of-plane (in-plane) along (perpendicular to) the easy-axis of the FGT electrodes. }
\label{Sample}
\end{figure*}

We observed spin signals in two very similar spin injection devices, sample A and sample B. Here we present the measurement results for sample A, while the measurements for sample B are summarized in the Supplementary Information (Fig.~S1, S2). A microscope image of sample A is shown in Fig.\,\ref{Sample}~(a). The device consists of a monolayer graphene channel with two ferromagnetic contacts on top, composed of an FGT/hBN structure. hBN, FGT and graphene were exfoliated onto p$^{++}$ doped silicon (Si) chips with a 90\,nm SiO$_2$ capping layer \footnote{hBN was grown by a high pressure technique. FGT was bought from HQ graphene and graphene was exfoliated from Flaggy Flakes natural graphite bought from NGS Naturgraphit GmbH}. However, the FGT flakes were exfoliated in a glovebox with an O$_2$  concentration below 0.1\,ppm. The widths of the two FGT flakes are 2.6\,\textmu m and 1.6\,\textmu m for sample A and 2.3\,\textmu m and 1.6\,\textmu m for sample B. The thicknesses of the injecting and detecting electrodes for sample A are 145\,nm and 85\,nm, respectively, while for sample B they are 66\,nm and 113\,nm. The distance between the two FGT flakes, which defines the length of the spin transport channel, is $d=5$\,\textmu m for sample A and 5.6\,\textmu m for sample B, measured between the centers of the flakes. The stack was assembled inside the glovebox on a p$^{++}$ doped Si chip with a 285\,nm thick layer of dielectric SiO$_2$ using a standard dry transfer technique employing polycarbonate \cite{Pizzocchero.2016}. The highly doped silicon is used as a global back gate. The graphene was then patterned into a Hall bar using electron beam lithography (EBL) and reactive ion etching (RIE). The width of the Hall bar is 3.5\,\textmu m. Subsequently, the contacts to the Hall bar and to the ferromagnetic electrodes were prepared using EBL and standard thermal evaporation of Ti(5\,nm)/Au(150\,nm). A schematic of the completed sample is shown in Fig.\,\ref{Sample}~(b). The layer sequence of the device thus consists of a monolayer of graphene on top of the SiO$_2$ substrate followed by a 0.9\,--\,1.3\,nm thick layer of hBN (measured with atomic force microscopy (AFM)) just below the FGT flakes, which are then covered by another hBN flake as a capping layer. The thin hBN flake acts as a tunnel barrier to ensure a good spin injection efficiency \cite{Schmidt.2000}. It is worth noting that the samples without a tunnel barrier did not show any spin signal.

All experiments were carried out in a cryostat capable of reaching temperatures as low as 1.5\,K, with the sample mounted on a rotating holder that allowed varying the angle between the sample and the applied external magnetic field. Spin injection experiments were performed in a standard non-local configuration (see Fig.\,\ref{Sample}(c)), with the charge current flowing between one of the FM contacts and a reference non-magnetic contact at the end of the mesa \cite{Johnson.1985}. The charge current flowing through the FGT/hBN/graphene structure generates a spin accumulation in graphene, which diffuses away from the junction in all directions (red shaded region in Fig.\,\ref{Sample}(c)). The spin accumulation can then be detected by the second FGT/hBN contact, placed at a distance $d$ from the injecting contact, outside the charge current path. The non-local voltage measured between the detector and the reference contact serves as a measure of the spin accumulation beneath the detector. The electronic measurements were carried out using a Yokogawa 7651 as the DC current source and a Keithley 2400 as a back gate voltage source. The measured non-local voltage was amplified by a FEMTO DLPVA-101 voltage amplifier that was connected to a SynkTek MCL1-540 multi-channel data acquisition system. Voltages at other voltage probes were measured with the data acquisition system alone. Since FGT has its magnetic easy-axis out-of-plane, the nonlocal spin valve experiments were all performed by sweeping the external magnetic field in this direction. For the Hanle measurements, the external magnetic field was swept in-plane, along the long axis of the spin contacts, perpendicular to the transport channel.\\

\section{Results and discussion}

\subsection{Electrical characterization}

\begin{figure*}
\includegraphics[keepaspectratio, width=\textwidth]{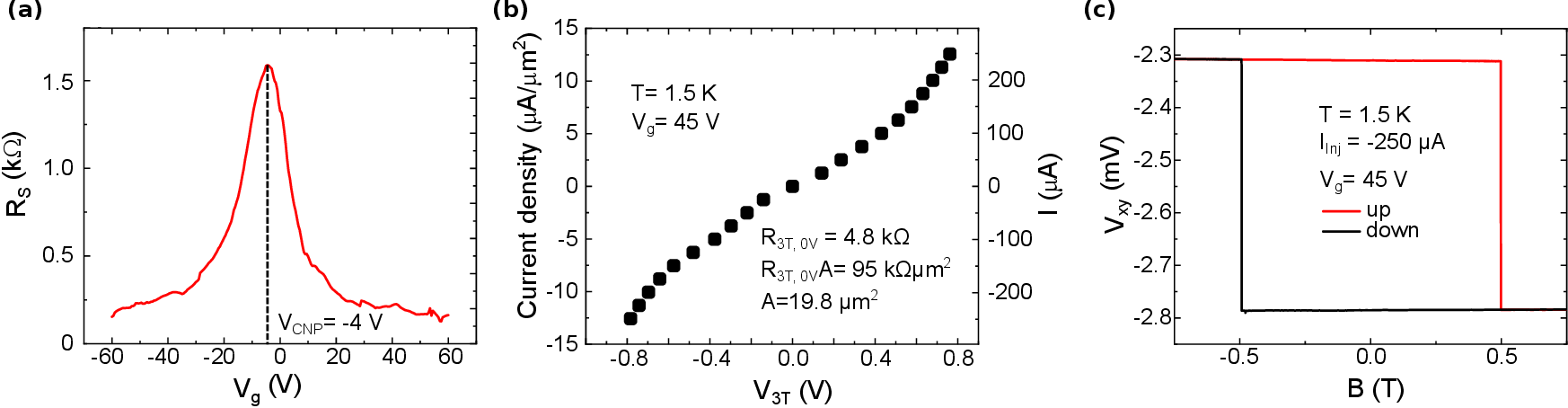}
\caption{Electronic characterization of the device. (a) Sheet resistance $R_\text{S}$ of the graphene channel in the reference area as a function of the back gate voltage. The measurements have been carried out at $T=1.5$\,K and with $I=50$\,\textmu A. (b) Characterization of the hBN tunnel barrier. The current density is plotted against the measured three-terminal voltage at the injecting electrode, which constitutes a voltage dropped across the tunnel barrier contact. The current density, shown on the left side, presents the current normalized by the area of the FGT/hBN contact. (c) Transversal voltage measured across the FGT flake, while sweeping the external magnetic field out-of-plane during the spin valve measurements shown in Fig.~\ref{SpinValve}(a). Red (black) line corresponds to the up (down) sweep. The observed switching corresponds to the switching of magnetization in the FGT flake.}
\label{Characterization}
\end{figure*}

Before describing the results of the spin measurements, we first discuss the electrical characterization of the device components. We characterized the graphene channel by measuring its sheet resistance $R_\text{S}$ as a function of the back gate voltage, determining the charge neutrality point (CNP) at $V_\text{g}=-4$ V, in the Hall bar section of the device (see Fig.\,\ref{Characterization}(a)). From this measurement, mobilities of up to 11000\,cm$^2$/Vs were extracted, consistent with the results of the Hall measurements (see Fig.\,S3 in the Supplementary Information). In Fig.\,\ref{Characterization}(b) we show the \mbox{$I$-$V$-curve} of the injection electrode, as a function of the three-terminal voltage. The zero-bias resistance-area product $R_\text{3T, 0V}A$ characterizes the tunnel barrier, according to Britnell $et\, al.$ \cite{Britnell.2012}. The measured $R_\text{3T,0V}A\sim 95$\,k$\Omega\cdot$\textmu m$^2$ corresponds to the hBN flake being two layers thick, which is consistent with the AFM measurements within the measurement accuracy. Furthermore, the switching behaviour of the injecting FGT electrode was monitored by measuring the transverse voltage across the FGT flake, while a constant current was sent from the injecting FGT electrode into the graphene (see Fig.\,\ref{Characterization}(c)). In a ferromagnet, the transverse voltage is composed of the regular and the anomalous Hall voltage, with the latter being proportional to the magnetization of the ferromagnet (\mbox{$R_{xy}=R_\text{RH}+R_\text{AH}=R_0 \mu_0 \cdot H + R_\text{S} \cdot  M$, \cite{Nagaosa.2010}}). Therefore, we can attribute a sharp step in the transverse anomalous Hall voltage to the abrupt switching of the magnetization in our injecting FGT electrode. This switching is consistent with the switching of the non-local voltage in spin-valve measurements, as described later.

\subsection{Non-local spin valve}

\begin{figure*}
\includegraphics[keepaspectratio, width=0.8\textwidth]{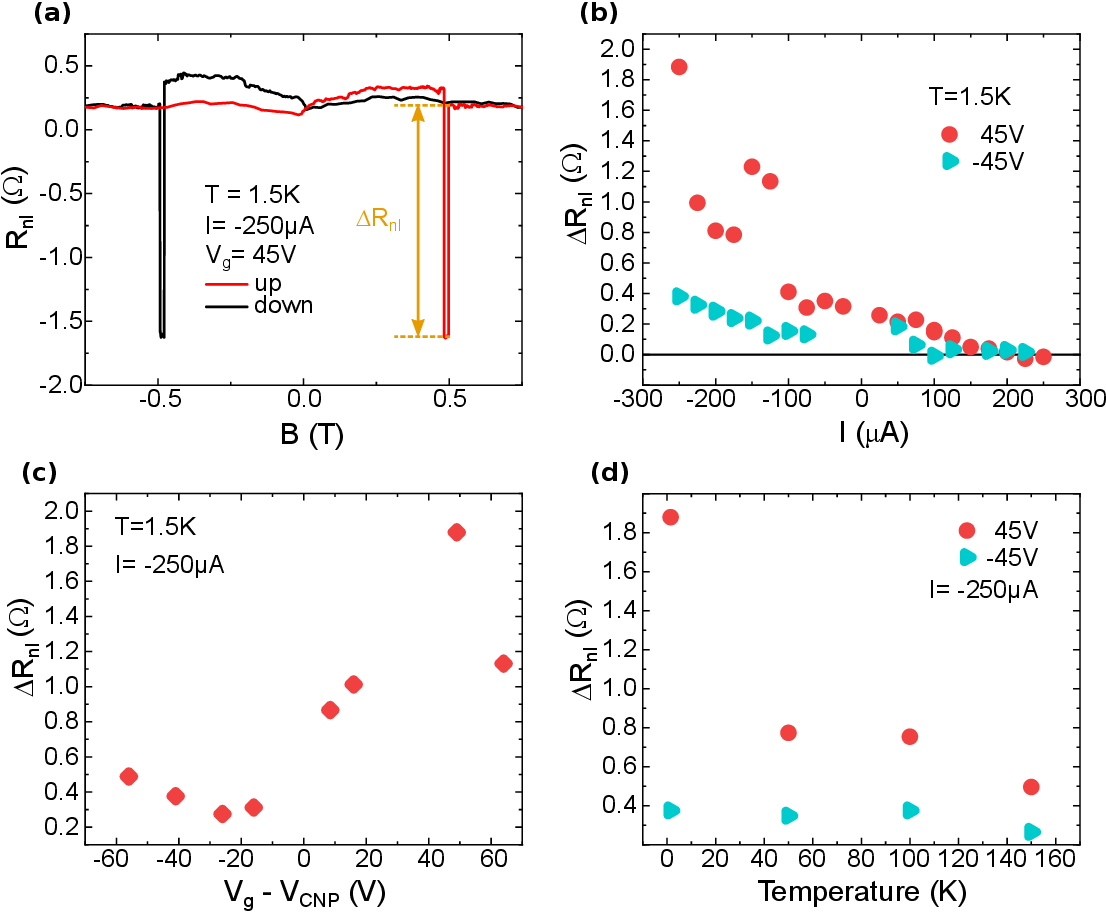}
\caption{(a) Non-local spin valve measurement at $T= 1.5$\,K, $V_{\text{g}}= 45$\,V and a current of $I=-250$\,\textmu A. The dip of the nonlocal signal at low magnetic fields is clearly discernible. In yellow, the height of the non-local signal $\Delta R_\text{nl}$ is shown. (b) Current, (c) back gate voltage and (d) temperature dependence of the non-local signal height in the electron ($V_{\text{g}}= 45$\,V) and hole ($V_{\text{g}}= -45$\,V) regime.}
\label{SpinValve}
\end{figure*}

Non-local spin-valve measurements are a standard way of detecting spin accumulation in lateral spin injection devices \cite{lou2007,Tombros.2007,Ciorga2009, Johnson.1985}. Here, a magnetic field is swept along the easy-axis of the spin electrodes, which in our case is oriented out-of-plane, and the non-local voltage $V_\text{nl}$ is measured at the detector, with a current flowing in the injector circuit. Changes in $V_\text{nl}$ are observed whenever the magnetization of one of the contacts switches, leading to a transition between parallel and anti-parallel magnetization configurations in the two spin aligning electrodes. In Fig.\,\ref{SpinValve}(a) we show a typical spin valve trace, where we plot $V_\text{nl}$ normalized by the injection current $I$ as a nonlocal resistance $R_\text{nl}=V_\text{nl}/I$. The amplitude of the switching $\Delta R_\text{nl}$ serves as a measure of the generated spin accumulation and is given by \cite{Fabian2007, Oltscher.2014}
\begin{equation}
\Delta R_\text{nl}=\frac{P_\text{inj}P_\text{det}R_\text{s}\lambda_\text{s}}{w}\exp\left(-\frac{d}{\lambda_\text{s}} \right).
\end{equation}
In the above equation, $\lambda_\text{s}$ is the spin diffusion length, $w$ is the width of the channel, and $P_\text{inj}$ and $P_\text{det}$ are the spin injection and detection efficiency, respectively. These efficiencies are defined as the spin polarization of the injected current directly underneath the given contact when the contact is used as an injector. Assuming the same interfaces at the injector and detector contacts and for low injection currents, one can take $P_\text{inj}\approx P_\text{det}= P$. In general, however, $P_\text{inj}$ can depend on the injection current, leading to a current dependence of the measured signal, as shown in Fig.\,\ref{SpinValve}(b). We plot here $\Delta R_\text{nl}$ measured at $T=1.5$\,K for gate voltages $V_\text{g}=45$\,V and $V_\text{g}=-45\,$V, corresponding to electron and hole transport in graphene, respectively. For both carrier polarities, $\Delta R_\text{nl}$ is higher for a negative bias, corresponding to injection of spin-polarized electrons from FGT into graphene or extraction of spin-polarized holes, respectively, and decreases almost monotonically, as the injection current is changed towards positive values. For very high positive currents at $T=1.5$\,K we even observe an inversion of the spin signal in the electron regime. This behaviour is typically driven by a change in the sign of $P_\text{inj}$ with bias, indicating an inversion of spin polarization around the Fermi level of the ferromagnetic material. This phenomenon has been observed previously in both conventional graphene spin valve devices \cite{Ringer.2018, Zhu.2018} and in III-V materials \cite{lou2007}. In recent experiments with Fe$_5$GeTe$_2$, it was shown that Fe$_5$GeTe$_2$ had an opposite spin polarization compared to that observed in Co electrodes for the entire range of bias currents used \cite{Zhao.2023}. In our experiments, the sign reversal of the spin valve signal is a result of the sign change in the tunneling density of states (TDOS) in our structure, as will be discussed later in more detail. 

In the entire range of bias currents, the signal is much stronger for electrons than for holes, which is confirmed by plotting $\Delta R_\text{nl}$ as a function of gate voltage, see Fig.\,\ref{SpinValve}(c). Additionally, it can be observed that $\Delta R_\text{nl}$ increases with the absolute value of $V_\text{g}$. Similar behaviour was also observed at higher temperatures, as can be seen in the supplementary Fig.\,S4. In Fig.\,\ref{SpinValve}(d) we plot $\Delta R_\text{nl}$ as a function of $T$ for $I=-250$\,\textmu A, showing a general trend of decreasing spin signal with increasing $T$. Whereas the current dependence of $\Delta R_\text{nl}$ can be linked to the bias dependence of $P_\text{inj}$, explaining its gate and temperature dependence requires information about gate and $T$-dependence of $P_\text{inj}$, $\lambda_\text{s}$, and $R_\text{s}$. To experimentally determine $\lambda_\text{s}$ and $P_\text{inj}$, we performed Hanle measurements, investigating spin precession in an external transversal magnetic field, which we will discuss in the next section. 

Apart from a clear spin-valve pattern, we also observed another feature in the spin-valve measurements, namely a dip in the non-local signal at low magnetic fields, see Fig.\,\ref{SpinValve}(a) and Fig. S5 in the Supplementary Information. Such a dip is typically associated with the presence of magnetic moments in a graphene channel, which introduce relaxation of spin currents through exchange coupling \cite{McCreary.2012, Birkner.2013}. Given that our samples were fabricated in an inert atmosphere and capped with hBN, and were not subjected to any hydrogenation \cite{McCreary.2012} or annealing \cite{Birkner.2013} processes, which are reported to induce magnetic moments, we cannot provide an explanation for the origin of these magnetic moments. However, the results of the Hanle measurements, discussed below, are also consistent with the presence of magnetic moments in the channel. 

\subsection{Hanle signal}

\begin{figure}
\includegraphics[keepaspectratio, width=0.5\textwidth]{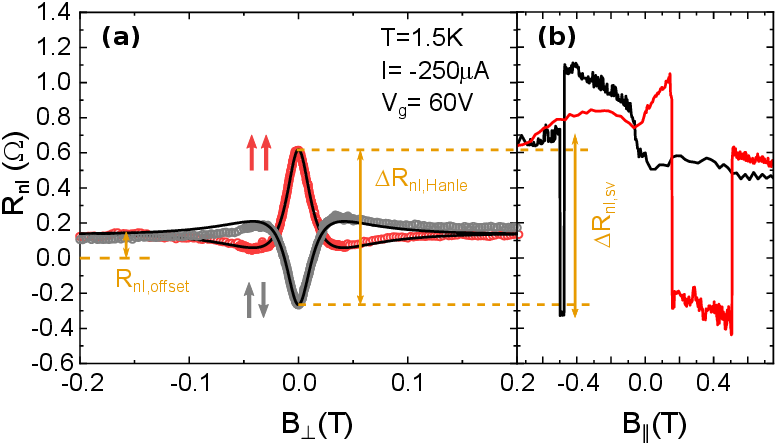}
\caption{ (a) Hanle signal for parallel (red) and anti-parallel (grey) magnetization alignment of the FGT electrodes. Fits to the measurements are shown as black curves. The non-local Hanle signal height at $B=0$\,T is $\Delta R_\text{nl, Hanle}= 0.88$\,$\Omega$. (b) Non-local spin valve measurement for the same measurement parameters. The spin valve height is $\Delta R_\text{nl, spin valve}=1.13$\,$\Omega$. The magnetic field is swept perpendicular and parallel to the magnetization direction of FGT in (a) and (b), respectively. }
\label{HanleGeneral}
\end{figure}

In Hanle measurements, the external magnetic field is applied transversely to the orientation of the injected spins, inducing their precession as they travel from the injector to the detector \cite{Tombros.2007}. As a result of diffusive motion and spin relaxation, the spins dephase and depolarize, which is reflected in the measured $V_\text{nl}$ \cite{Fabian2007}. In Fig.\,\ref{HanleGeneral}(a) we plot a Hanle signal for the injection current $I=-250$\,\textmu A, at a temperature $T=1.5$\,K and a back-gate voltage of $V_{\text{g}}=60$\,V for the anti-parallel (grey) and parallel (red) magnetization configuration of the two ferromagnetic electrodes of sample A. The similar plot for sample B can be seen in Fig.\,S1(d). Since the spins injected from FGT are polarized out of plane, we applied an external in-plane magnetic field, parallel to the long axis of FGT flakes. The difference of the signal measured for parallel and anti-parallel sweeps at $B=0$\,T gives \mbox{$\Delta R_{\text{nl,Hanle}}= 0.88$\,$\Omega$}, which is slightly lower than the corresponding spin valve signal $\Delta R_{\text{nl, sv}}= 1.13$\,$\Omega$ (see Fig.\,\ref{HanleGeneral}(b)). The small discrepancy between the Hanle and spin valve signals may be attributed to the presence of magnetic moments, which reduce the spin signal at low magnetic fields, thereby reducing the height of the Hanle curve. This observation is consistent with the findings of the non-local spin valve measurements, which also indicated the presence of magnetic moments. It is noteworthy that these magnetic moments are believed to be extrinsic and not associated with the FGT electrodes. However, we cannot rule out a hysteresis of the signal due to the measurement procedure, as we first recorded the spin valves for all currents and back gate voltages and afterwards performed the Hanle sweeps. The small hysteresis with respect to the current or gate voltage cannot be excluded and could potentially lead to a smaller signal in the Hanle curves. There is also a small asymmetry between the signal in parallel and anti-parallel configuration, which we cannot account for at the moment.

\begin{figure*}
\includegraphics[keepaspectratio, width=\textwidth]{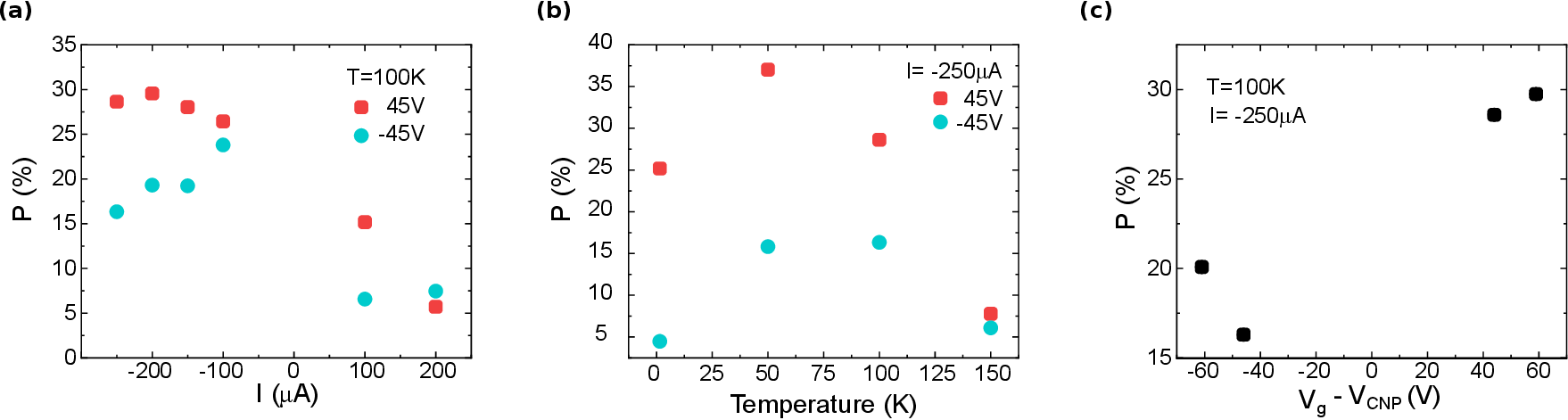}
\caption{(a) Current, (b) temperature and (c) gate voltage dependence of the spin injection efficiency $P$ obtained from fits to the Hanle curves. Error bars, determined as the fitting errors, are in most cases smaller than the size of the symbols.}
\label{HanleResults}
\end{figure*}

At finite transverse fields, we clearly observe the oscillations of the signal as a result of spin precession and simultaneous decay of the signal as a result of spin dephasing. At a sufficiently large magnetic field $B\gtrsim 0.2$ T, spins depolarize through dephasing, the spin signal approaches zero, and the measured non-local resistance $R_{\text{nl, offset}}$ constitutes the non-local baseline resistance\cite{Bakker2010,Johnson2007}. The solid lines in Fig.\,\ref{HanleGeneral}(a) are fitting curves based on the steady-state solution of the spin drift-diffusion equation \cite{Fabian2007} $\frac{\partial{\bm{\mu_\text{s}}}}{\partial{t}}=\bm{\mu_\text{s}}\times \bm{\omega_L}+D_\text{s}\nabla^2\bm{\mu_\text{s}}-\frac{\bm{\mu_\text{s}}}{\tau_\text{s}}$, with the boundary condition at the injector $e^2D_\text{s}\nu(E_\text{F})\nabla \bm{\mu_\text{s}}=P_\text{inj}\bm{j}$. In the above equations, $\mu_\text{s}$ indicates spin accumulation generated by the injection current density $\bm{j}$, which at the detector is measured as a non-local voltage $V_\text{nl}=-P_\text{det}\mu_\text{s}(d)$, $D_\text{s}$ is the spin diffusion constant, $\tau_\text{s}$ spin relaxation time, $\bm{\omega_L}=\frac{g^*\mu_\text{B}\bm B}{\hbar}$ is the Larmor frequency at the external magnetic field $\bm B$, with $g^*=2$ being the Land\'e factor, $\nu(E_\text{F})$ is the density of states at the Fermi level, $\mu_\text{B}$ is the Bohr's magneton and $\hbar$ the reduced Planck's constant. From the fitting curves, we obtain the values of $P$, $D_\text{s}$ and $\tau_\text{s}$, with the latter two giving the spin diffusion length $\lambda_\text{s}=\sqrt{D_\text{s}\tau_\text{s}}$. The extracted value of $P$ is $P=\sqrt{P_\text{inj} P_\text{det}}$. To minimise errors in fitting these three variables, we first fitted the normalised Hanle data \mbox{$(R_{\text{nl, B}}-R_{\text{nl, offset}})/(R_{\text{nl, 0T}} - R_{\text{nl, offset}})$} to extract $\tau_{\text{s}}$ and $D_\text{s}$ from the shape of the curves and then we fitted the raw data with the extracted values from the normalised fits. Therefore, $P$ was the only variable in the second fit. As can be seen in Fig.\,\ref{HanleGeneral}(a) the fits (shown as a black line) match the experimental data quite well. Fitting the parallel Hanle curve gives $\tau_\text{s}=0.447$\,ns, $D_\text{s}= 0.0210$\,$\frac{\text{m}^2}{\text{s}}$, and $P=18.4$\,\% , whereas we obtain $\tau_\text{s}=0.415$\,ns, $D_\text{s}=0.0199$\,$\frac{\text{m}^2}{\text{s}}$, and $P=18.3$\,\% in the anti-parallel configuration. 

We performed Hanle measurements in the parallel configuration for different injection currents, back gate voltages and at different temperatures. The full set of results for sample A and B can be found in the supplementary Fig.\,S6 and Fig.\,S2, respectively. The fitting results for $P$ of sample A are summarized in Fig.\,\ref{HanleResults}. In the following section, we discuss in more detail the obtained results.

\subsection{Discussion}

As can be seen in Fig.\,\ref{HanleResults}, we have obtained a fairly high injection efficiency, reaching up to 40\%, which is significantly higher than that reported for structures without tunnel barriers \cite{He.2023}. However, this is a low estimate of $P_\text{inj}$. When linearly extrapolating $P_\text{det}(1.5\text{\,K})$=$P(0\text{\,\textmu A, 1.5\,K})$ to be $\approx 17$\,\%, the spin injection efficiency is estimated to be $P_\text{inj}(-200\text{\,\textmu A}, 45\text{\,V}, 1.5\text{\,K})=93$\,\%. Consistent with the spin valve signal $\Delta R_{\text{nl}}$, $P$ is larger for the negative back gate voltages, i.e., in the electron regime, as shown in Fig.\,\ref{HanleResults}(c), and for negative injection currents, i.e., for the case of electron injection. $P$ decreases, while sweeping the injection current from negative to positive values as illustrated in Fig.\,\ref{HanleResults}(a). As bias affects only the injector, the decrease in $P$ with current is attributed to a decrease of $P_\text{inj}$. $P$ also decreases with increasing temperature for $T\geq 50$\,K, although at $T=1.5$\,K $P$ is lower than at $T=50$\,K, both in the electron and hole regime, as shown in Fig.\,\ref{HanleResults}(b).

In order to properly interpret the current dependence of the spin injection efficiency (as shown in Fig.~\ref{HanleResults}(a)), it is helpful to have some knowledge about the spin polarization of Fe$_3$GeTe$_2$. To this end, we performed DFT calculations of the electronic band structure of the bulk FGT (see the Supplementary Information IV for details), including the spin-resolved density of states (DOS). A measure for the degree of spin polarization of the injected current is the tunneling density of states (TDOS), which is defined via the product of DOS and the velocity of the Bloch bands~\cite{Mazin1999:PRL}. It should be noted that this calculation does not take into account any properties of the interface, barrier or second contact. Based on spin-resolved DOS, $N_{\uparrow / \downarrow}$, and Bloch band velocities in the $z$-direction (perpendicular to the Fe$_3$GeTe$_2$ layers), $v_z$, we calculate the DOS spin polarization $P_N$ of the bulk FGT and the TDOS spin polarization $P_{Nv}$ and $P_{Nv^2}$ as follows~\cite{Mazin1999:PRL}:

\begin{equation}
P_N = \frac{N_{\uparrow}-N_{\downarrow}}{N_{\uparrow}+N_{\downarrow}},\\
\end{equation}
\begin{equation}
    P_{Nv} = \frac{\langle Nv_z \rangle_{\uparrow}-\langle Nv_z \rangle_{\downarrow}}{\langle Nv_z \rangle_{\uparrow}+\langle Nv_z \rangle_{\downarrow}}\\
\end{equation}
\begin{equation}
    P_{Nv^2} = \frac{\langle Nv_z^2 \rangle_{\uparrow}-\langle Nv_z^2 \rangle_{\downarrow}}{\langle Nv_z^2 \rangle_{\uparrow}+\langle Nv_z^2 \rangle_{\downarrow}}.
\end{equation}

\begin{figure*}
\includegraphics[keepaspectratio, width=0.7\textwidth]{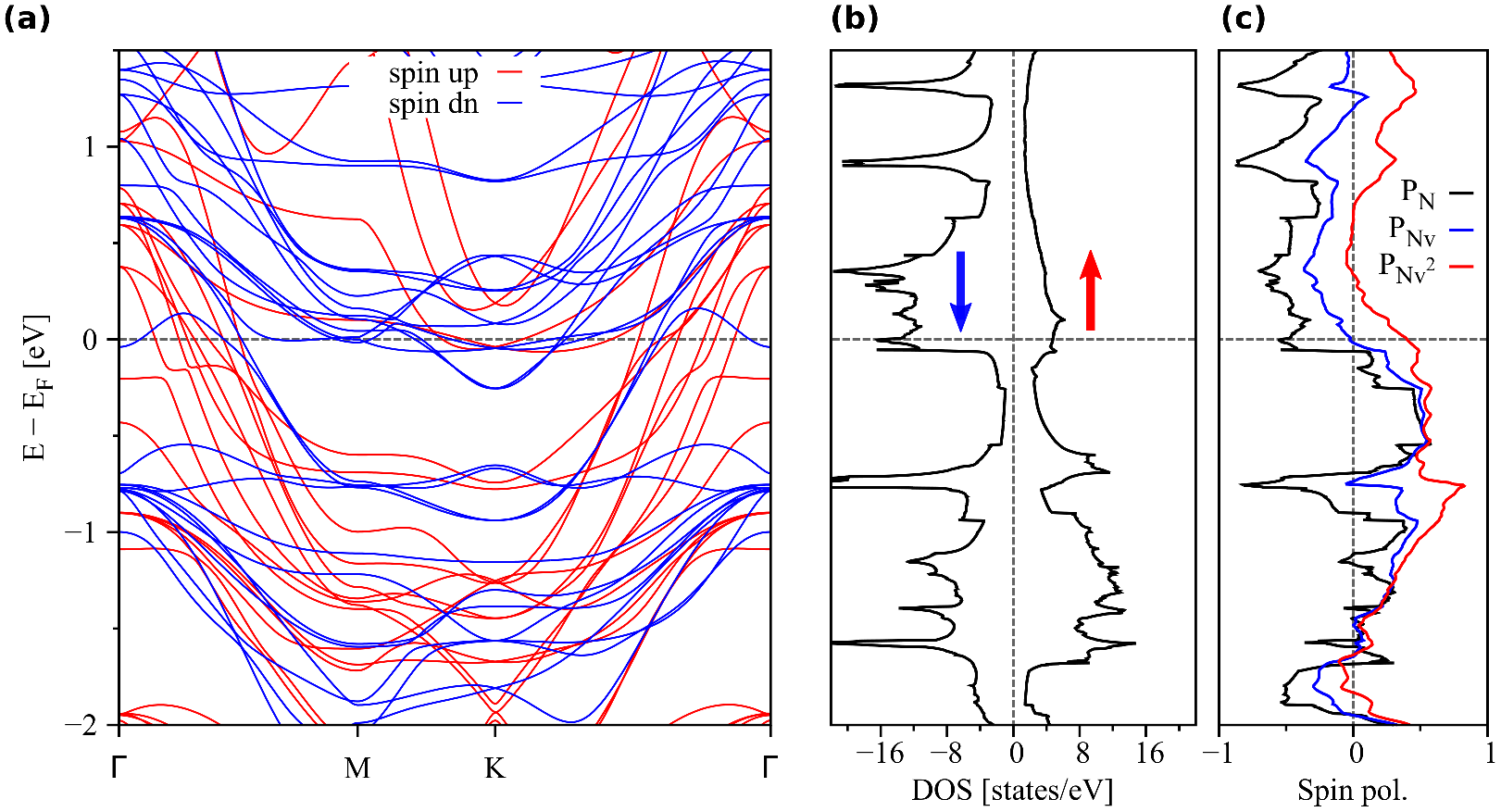}
\caption{(a) DFT-calculated band structure of bulk Fe$_3$GeTe$_2$. Red (blue) lines correspond to spin up (down). (b) The corresponding spin-resolved density of states. Negative (positive) values for the DOS reflect a spin down (up) polarization, as indicated by the blue (red) arrow. (c) Calculated results for the spin polarization $P_N$, $P_{Nv}$ and $P_{Nv^2}$.}
\label{FIG:TDOS}
\end{figure*}

In Fig.\,\ref{FIG:TDOS}~(a) and (b) we show the calculated FGT band structure and its spin-resolved DOS, respectively. Additionally, the spin polarization $P_N$, together with the TDOS $P_{Nv}$ and $P_{Nv^2}$ are shown in Fig.~\ref{FIG:TDOS}(c). We note that at the Fermi level $E_F$ and at higher energies, the DOS is highly spin-polarized with the majority of spin-down states. However, slightly below $E_F$, the DOS decreases, particularly for spin-down states. Both, $P_N$ and $P_{Nv}$, change sign below $E_F$. In contrast to this, $P_{Nv^2}$ tends to stay positive in the close vicinity of $E_F$, indicating that the current is dominated by the spin-up charge carriers. However, the degree of spin polarization of $P_{Nv^2}$ decreases, towards larger energies. In the experiment, we tune the alignment of the Fermi-level of graphene and FGT by changing the bias across the junction and we note that the calculated decrease of $P_{Nv^2}$ towards larger energies is very consistent with the measured decreasing spin injection efficiency towards larger positive currents, shown in Fig.\,\ref{HanleResults}(a).

In order to obtain a comprehensive understanding of the tunneling, it is necessary to calculate the coherent tunneling for the entire FGT/hBN/hBN/graphene structure. This calculation requires precise knowledge of the band structure and the exact twist angles of each layer. However, as we lack access to this structural information, and given the focus of this paper on the experimental realization of efficient spin injection and detection in all van der Waals heterostructures, these calculations cannot be performed and are beyond the scope of the presented work. Nevertheless, a change of sign at or near the $E_F$ is evident for all calculated spin polarizations $P_N$, and $P_{Nv}$, and $P_{Nv^2}$ of FGT, which might provide an explanation for the current dependence of the non-local signal height (see Fig.~\ref{SpinValve}~(b)) and the spin injection efficiency (see Fig.~\ref{HanleResults}~(a)).

Let us now discuss the obtained spin transport parameters. The extracted values for $\tau_\text{s}$ are in the range from $\sim$300\,ps to $\sim$600\,ps and $D_\text{s}$ spans from $\sim$0.004\,m$^2$/s to $\sim$0.04\,m$^2$/s. There is no clear dependence of both variables on current and on gate voltage. However, fit results of $\tau_\text{s}$ and $D_\text{s}$ both suggest larger values for negative than for positive back gate voltages, i.e. in the hole conduction regime (see Fig.\,S6). Furthermore, a small dependence of $\tau_{\text{s}}$ and $D_\text{s}$ on temperature is observed. Whereas the values extracted for $T=1.5$\,K are larger than at $T=50$\,K, for \mbox{$T\geq 50$\,K} the spin relaxation time increases from $\tau_{\text{s}}$= 0.3\,ns at $T=50$\,K to 0.5\,ns at $T=150$\,K in the electron regime and a similar effect can also be seen in the hole regime (see Fig.\,S6). Also $D_\text{s}$ increases with temperature in a similar way as $\tau_\text{s}$, so the calculated spin diffusion length $\lambda_\text{s}=\sqrt{D_\text{s}\tau_\text{s}}$ doubles from 1.5\,\textmu m to 3.1\,\textmu m in the electron regime and increases from 2.24\,\textmu m to 2.78\,\textmu m in the hole regime. 

Surprisingly, the extracted values of $D_\text{s}$ are significantly lower than the values of the charge diffusion constant $D_\text{c}$ at the same temperatures and gate voltages, as obtained from transport measurements, which are in the range $0.08-0.12$\,m\textsuperscript{2}/s (see supplementary Fig.\,S8). This discrepancy between the charge and spin diffusion constants and the temperature dependence of $\tau_\text{s}$ could be explained by the presence of magnetic moments, which would be consistent with the spin valve measurements. Resonant scattering at magnetic impurities introduces a temperature-dependent scattering rate \cite{Kochan.2014} and results in narrower Hanle curves due to the additional exchange field \cite{McCreary.2012}. This exchange field can be taken into account in the Hanle curve fitting, taking a larger effective g-factor $g^*>2$. During the above-described fitting of the Hanle curves, a constant g-factor of $g^*=2$ was assumed, which in the presence of magnetic moments results in incorrect values of $D_\text{s}$. To correct for this, we performed an alternative fitting, where we fixed $D_\text{s}=D_\text{c}$ and extracted from the fitting the effective g-factor. However, this resulted in very large values of the effective g-factor, reaching as high as $g^*_\text{eff}=23$. This would indicate the presence of a substantial exchange field or a significant amount of magnetic moments in the graphene channel, whose origin is unknown to us. 

Another explanation for the peculiar temperature dependence of $\tau_{\text{s}}$ and $D_\text{s}$, and the low values of $D_\text{s}$, could be provided by the assumption, that under contacts both $\tau_{\text{s}}$ and $D_\text{s}$ are strongly suppressed because of the influence of the ferromagnetic FGT. As the fitting was performed assuming uniform $\tau_{\text{s}}$ and $D_\text{s}$ throughout the channel, the extracted values of both parameters could be underestimated. As with increasing temperature the magnetization of FGT decreases (see Fig.\,S9), so does its possible detrimental effect on the spin dynamics in graphene. As a result, the extracted $\tau_{\text{s}}$ and $D_\text{s}$ would increase. In order to investigate a potential magnetic proximity effect \cite{Zollner.2016, Zollner.2022} at the FGT/hBN/graphene interface, we performed density functional theory (DFT) calculations with a two-layer hBN tunnel barrier (see Supplementary Information IV for details). In the calculated band structure of the heterostructure, the Dirac states of graphene remain spin-degenerate, and no magnetic moments are induced. Consequently, a proximity effect in graphene due to the FGT can be ruled out and cannot explain the discrepancy of $D_\text{c}$ and $D_\text{s}$.

\section{Conclusion}

In conclusion, we report on efficient electrical spin transport and spin precession in an all-van der Waals 2D device. Non-local signals are as large as \mbox{$\Delta R_\text{nl} \approx 1.9$\,$\Omega$}, showing a strong current dependence, and even leading to the inversion of the signal. The clear Hanle signal allowed for a full gate-, temperature-, and current-dependent characterization of the spin transport properties. A low estimate of the spin injection efficiencies results in \mbox{$P(-200$\,\textmu A$, 45$\,V$, 1.5$\,K$) = 40$\,$\%$}. The observed bias dependence of the spin injection efficiency, and the inversion of the spin valve signal are consistent with the calculated tunneling density of states. The presence of a small dip in the nonlocal spin valve measurements as well as the discrepancy between $D_\text{s}$ and $D_\text{c}$ suggest the presence of magnetic moments, whose origin, however, remains unknown.\\

\section*{Acknowledgements}
J.B., J. E., K. Z., L. C. and J. F. gratefully acknowledge support from the Deutsche Forschungsgemeinschaft (DFG, German Research Foundation) SFB 1277 (Project No. 314695032, sub-Project B07, A09), SPP 2244 (Project No. 443416183), the EU Graphene Flagship project 2DSPIN-TECH (Project No. 101135853), and FLAGERA project 2DSOTECH. K.W. and T.T. acknowledge support from the Elemental Strategy Initiative conducted by the MEXT, Japan (Grant Number JPMXP0112101001) and JSPS KAKENHI (Grant Numbers 19H05790 and JP20H00354). M. C. acknowledges support by the National Science Centre, Poland, project no. 2022/45/B/ST5/04292 of OPUS-23 call. We would also like to thank C. Strunk for facilitating access to the reactive ion etching system.

\bibliography{bibliography}
\ifarXiv


\section*{Supplementary material}

# Supplementary Information: Spin injection and detection in all-van der Waals 2D devices

## I. SUMMARY OF MEASUREMENTS ON SAMPLE B

Fig. S1 shows the summary of results obtain from measurements performed on sample B. In Fig. S1(a) the Dirac measurement with a charge neutrality point at $V_{\text{CNP}} = 11\,\text{V}$ is presented. From this measurement a mean mobility of $\mu = 9500\,\text{cm}^2/\text{Vs}$ is extracted, consistent with that of sample A. Fig. S1(b) shows the $I - V_{3\text{T}}$ characteristics of the injecting electrode of sample B, where a zero-bias resistance of $R_{3\text{T},\,0\text{V}} \approx 700\,\Omega$ is extracted from the curve. The non-local spin valve signal measured at $T = 70\,\text{K}$ is shown in Fig. S1(c). Interestingly, the sign of the non-local spin valve signal $\Delta R_{\text{nl}}$ is inverted when compared to the signal from sample A, as can be seen also in (e)–(h). Furthermore, the same sign of $\Delta R_{\text{nl}}$ was observed as for sample A, when the injecting and detecting FGT electrodes were interchanged. In Fig. S1(d) we show an example of Hanle signal observed for sample B, both in the parallel and anti-parallel magnetization configurations. The signal in both configurations is reversed when compared to that of sample A, consistent with the results of spin valve measurements. The values of all three parameters, $P$, $\tau_{\text{s}}$, and $D_{\text{s}}$, extracted from Hanle fits for sample B are shown in Fig. S2.

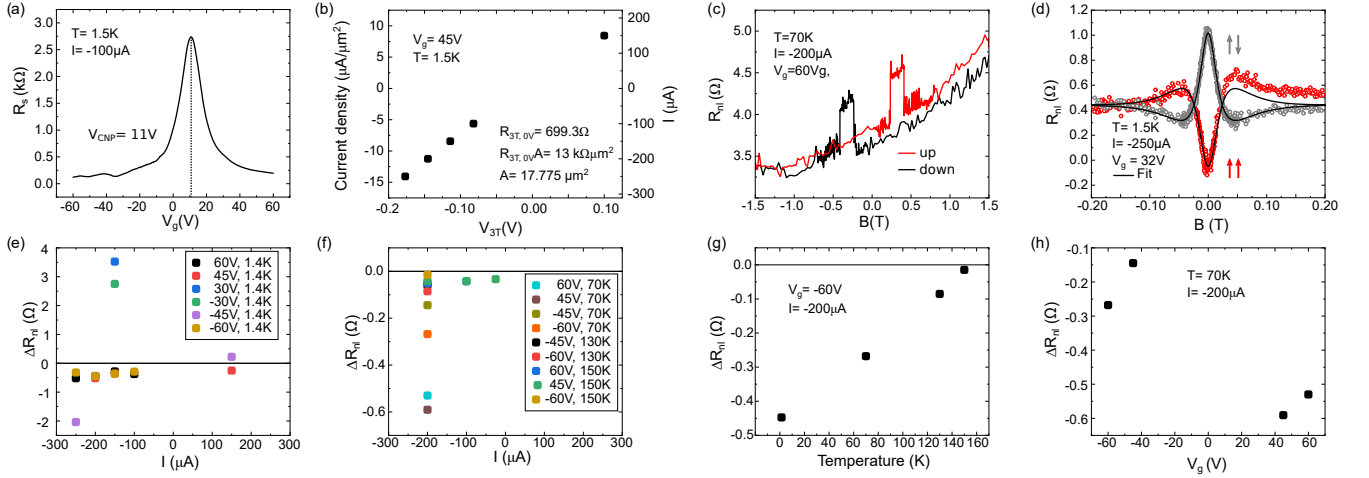

FIG. S1. Results of measurements on sample B. a) Sheet resistance of the graphene channel in the reference area as a function of the back gate voltage. b) Characterization of the hBN tunnel barrier. c) Typical non-local spin valve signal. d) Hanle signal for (anti-)parallel magnetization (shown in red (grey)) of the FGT electrodes, showing an inversion of the signal. Fits to the measurements are shown as black curves. e) and f) Current dependence of the non-local spin-valve signal height at different back gate voltages and temperatures. g) Temperature dependence of the non-local spin valve signal height. h) Back gate voltage dependence of the non-local spin valve height.

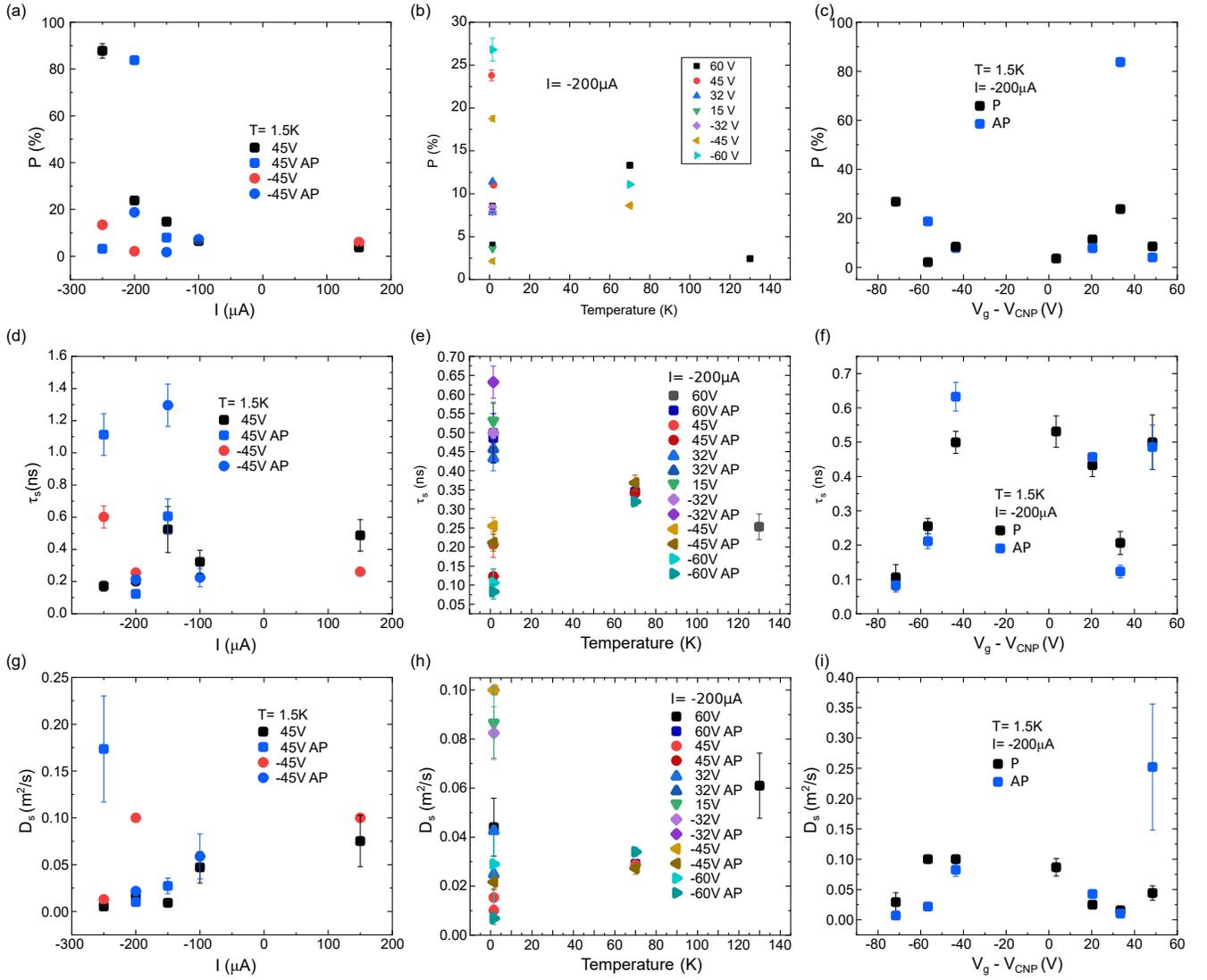

FIG. S2. Parameters extracted from the Hanle fits for sample B. Current (a), d) and g)), temperature (b), e) and h)) and gate voltage (c), f) and i)) dependence of the spin injection efficiency $P$, spin diffusion time $\tau_s$ and spin diffusion constant $D_s$. Error bars, determined as the fitting errors, are in most cases smaller than the size of the symbols.

## II. ADDITIONAL RESULTS FROM SAMPLE A

### A. Hall measurements of Sample A

Fig. S3 presents additional Hall measurements to verify the charge carrier densities determined via the Dirac cone measurements. From the Hall measurements, an average hole mobility of $\mu_{h,\,Hall} \approx 11000\,\text{cm}^2/\text{Vs}$, and an electron mobility of $\mu_{e,\,Hall} \approx 7500\,\text{cm}^2/\text{Vs}$ is obtained. These values are in good agreement with the Dirac cone measurements.

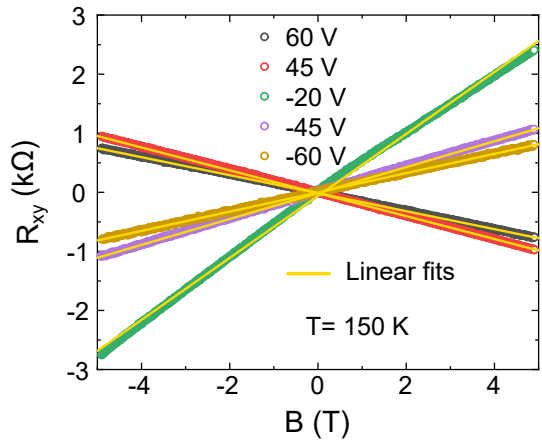

FIG. S3. Hall measurements at specific back gate voltages. The charge carrier density is determined by the inverse slope of the linear fits.

### B. Back gate voltage dependence

Fig. S4 shows the back gate voltage dependence of the non-local spin valve signal height $\Delta R_{nl}$ at elevated temperatures of $T = 50\,\text{K}$, $100\,\text{K}$, and $150\,\text{K}$. All measurements show the same qualitative behavior. i.e., larger spin signals in the electron regime than in the hole regime.

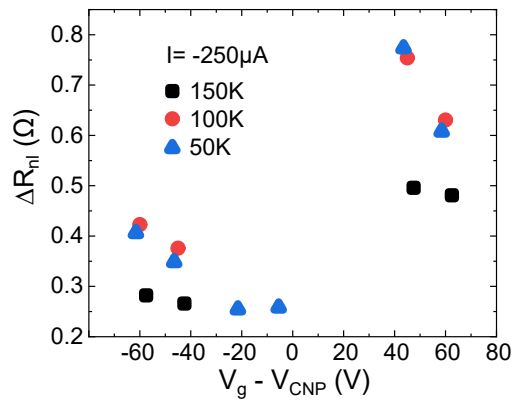

FIG. S4. Sample A: Non-local spin valve heights vs back gate voltage at higher temperatures.

### C. Signatures of magnetic moments

We observed dips at low magnetic fields in the nonlocal spin valve signals as can be seen in Fig. S5 at $T = 100\,\text{K}$. Such a dip was observed for all investigated temperatures, backgate voltages, and applied biases, for both up- and

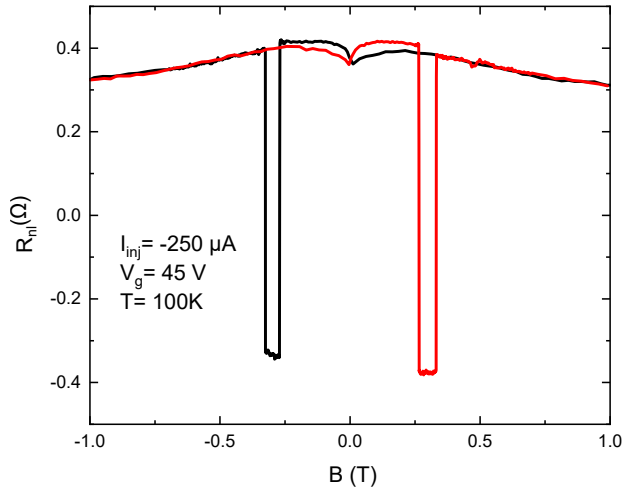

FIG. S5. Sample A: Non-local spin valve measurement at $T = 100\,\mathrm{K}$, $I = -250\,\mathrm{\mu A}$ and $V_g = 45\,\mathrm{V}$. At low magnetic fields, a clear drop in the non-local signal is observed, a feature typically related to the presence of magnetic moments in the graphene channel.

down-sweeps of the external magnetic field, being more pronounced in some curves than in others. This feature was observed in other graphene spin valve devices and is typically associated with the presence of magnetic moments on top of the graphene channel, that introduce spin relaxation through exchange coupling.

### D. Summary of Hanle results

In Fig. S6 we show the current, temperature, and gate voltage dependence of the spin relaxation time $\tau_s$ and the spin diffusion constant $D_s$. The values obtained for $\tau_s$, range from $0.315\,\mathrm{ns}$ to $0.383\,\mathrm{ns}$ at $V_g = 45\,\mathrm{V}$ and from $0.314\,\mathrm{ns}$ to $0.423\,\mathrm{ns}$ at $V_g = -45\,\mathrm{V}$, with no clear correlation with the current. The error bars indicate the errors from the fits and are mostly obscured by the dots. As the spin signal decreases with increasingly more positive currents, the signal-to-noise ratio decreases, resulting in very noisy Hanle signals at $I = 200\,\mathrm{\mu A}$. Consequently, the accuracy of the fit parameters is also reduced, resulting in large error bars for values at positive currents. The derived values for $\tau_s$ are in a range from $0.270\,\mathrm{ns}$ to $0.440\,\mathrm{ns}$ with no discernible dependence on the gate voltage. This indicates that the spin relaxation time $\tau_s$ is independent of the charge carrier type, and thus is identical in the electron and hole conduction regimes of graphene. Interestingly, an increasing tendency of $\tau_s$ for temperatures $T \geq 50\,\mathrm{K}$ can be observed for both applied gate voltages. While the values at $T = 1.5\,\mathrm{K}$ are larger than at $T = 50\,\mathrm{K}$, the spin relaxation time increases from $\tau_s = 0.3$ to $0.5\,\mathrm{ns}$ in the electron regime for $T \geq 50\,\mathrm{K}$. Such a temperature dependence of $\tau_s$ is consistent with spin relaxation caused by resonant scattering at magnetic impurities [1]. In this scenario, a high spin relaxation rate $1/\tau_s$ is calculated for $70\,\mathrm{K}$, an intermediate rate for $4\,\mathrm{K}$, and a low rate at $300\,\mathrm{K}$. This results in a temperature dependence of $\tau_s$ as observed in Fig. S6(b). As signatures of magnetic moments have also been observed in the non-local spin valve measurements, this behaviour may be attributed to these moments. Nevertheless, curves with comparable shapes can yield a variation in $\tau_s$ varying from $0.27\,\mathrm{ns}$ to $0.44\,\mathrm{ns}$. Therefore, the scattered values of $\tau_s = 0.3\,\mathrm{ns}$ to $0.5\,\mathrm{ns}$ when increasing the temperature from $T = 50\,\mathrm{K}$ to $150\,\mathrm{K}$ may also be attributed to the significant variation in the fit results.

The spin diffusion constant varies from $D_s = 0.0086$ to $0.012\,\mathrm{m^2 s^{-1}}$ with two outliers at $D_s(200\,\mathrm{\mu A}, 45\,\mathrm{V}) = 0.0046\,\mathrm{m^2 s^{-1}}$ and $D_s(100\,\mathrm{\mu A}, -45\,\mathrm{V}) = 0.0162\,\mathrm{m^2 s^{-1}}$. The spin diffusion constant $D_s$ does not show any systematic dependence on the current $I$. Upon investigating the gate dependence at $T = 100\,\mathrm{K}$, the spin diffusion constant was observed to vary from $D_s = 0.0043$ to $0.016\,\mathrm{m^2 s^{-1}}$, exhibiting no discernible correlation. The temperature dependence of $D_s$ reveals an increasing tendency from $D_s = 0.0075\,\mathrm{m^2 s^{-1}}$ to $0.0185\,\mathrm{m^2 s^{-1}}$ for $T \geq 50\,\mathrm{K}$, analogous to $\tau_s$. It remains unclear whether this observed increase is due to the large variation in the fitting results obtained for curves of similar shapes, or if it is a genuine increase of $D_s$ with temperature.

The largest values of $P$ where found for $T = 1.5\,\mathrm{K}$, $I = -200\,\mathrm{\mu A}$, and $V_g = 45\,\mathrm{V}$, as shown in Fig. S7. Linearly interpolating, one obtains $P_{\mathrm{det}}(1.5\,\mathrm{K}, 45\,\mathrm{V}) = P(0\,\mathrm{\mu A}, 1.5\,\mathrm{K}, 45\,\mathrm{V}) = 17\,\%$. When assuming $P_{\mathrm{det}}$ to be fixed at this

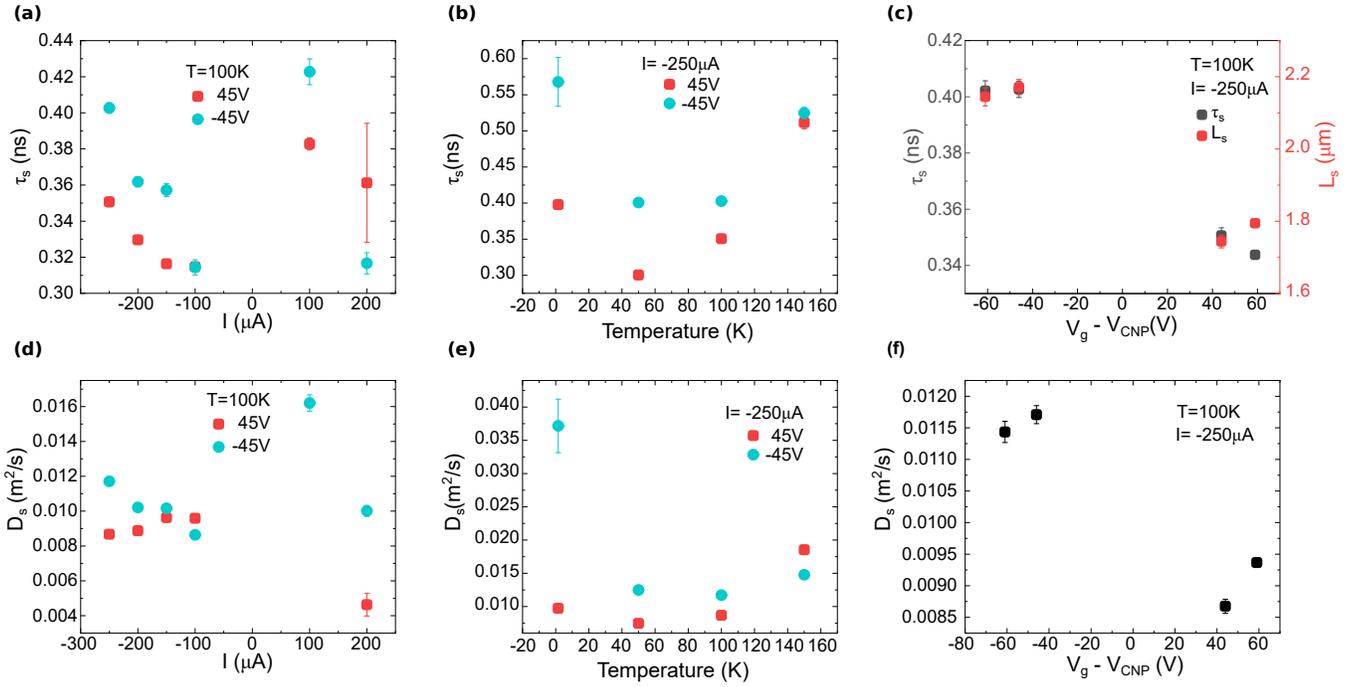

FIG. S6. Current (a) and d)), temperature (b) and e)) and gate voltage (c) and f)) dependence of the Hanle fitting results for $\tau_s$ and $D_s$ of sample A. Error bars, determined as the fitting errors, are in most cases smaller than the size of the symbols.

value, as it is not influenced by the current, a maximum value of $P_{\text{inj}}(200\,\mu\text{A}, 1.5\,\text{K}, 45\,\text{V}) = 93\,\%$ can be extracted.

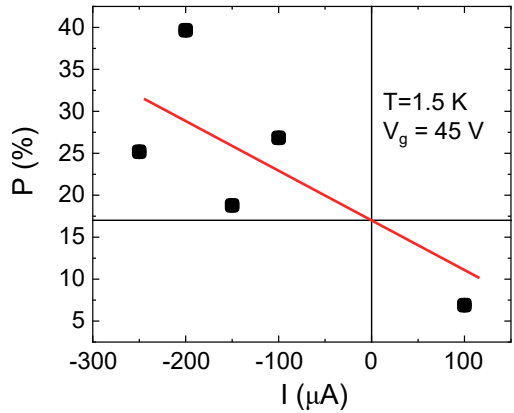

FIG. S7. Current dependence of the spin injection efficiency $P$ at $T = 1.5\,\text{K}$ at $V_g = 45\,\text{V}$. The linear extraction shown as a red line intercepts the $y$-axis at $P \approx 17\,\%$, which is also indicated by the black horizontal line.

### E. Comparison of charge and spin diffusion constants

The charge and spin diffusion constants, obtained from Dirac measurements and Hanle fits, respectively, are compared in Fig. S8. Generally, obtained values of $D_S$ are much smaller, than corresponding values of $D_c$.

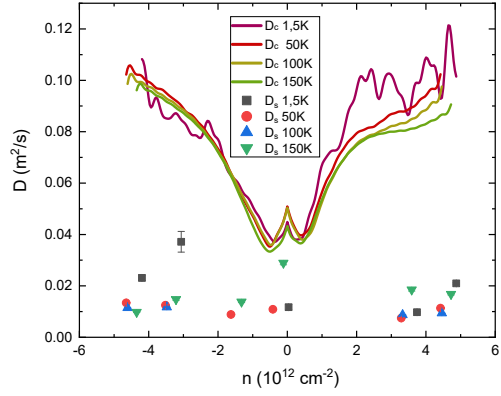

FIG. S8. Comparison of the charge and spin diffusion constants, $D_\text{c}$ and $D_\text{s}$, respectively, at various temperatures. The charge diffusion constants were obtained from the Dirac measurements while $D_\text{s}$ is obtained from Hanle fits.

## III. MAGNETIZATION OF FGT

We also studied the magnetization of single FGT flakes via the anomalous Hall effect. In Fig. S9(a), an optical micrograph of such a flake encapsulated with hBN is shown. The FGT is contacted via eight Cr/Au contacts. Fig. S9(b) and (c) show the anomalous Hall sweeps at various temperatures up to 200 K. The height of the anomalous Hall sweeps scales with the magnetization when assuming a constant anomalous Hall resistivity. From Fig. S9(d) a Curie-temperature of $T_\text{C} = 212.8$ K is extracted.

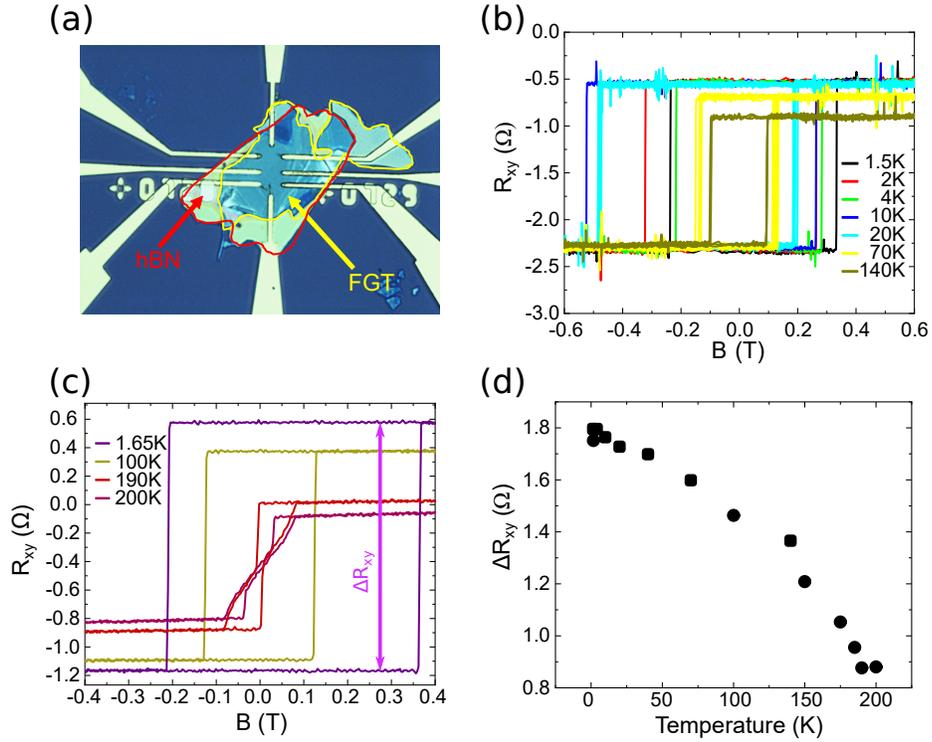

FIG. S9. a) Optical micrograph of an encapsulated FGT flake with a thickness of 66 nm. b) and c) show the anomalous Hall effect of the FGT flake at different temperatures, with clear switching of its magnetization. d) The height of the anomalous Hall loops is plotted versus the temperature, showing a clear decrease in the magnetization of the FGT flake.

## IV. DFT CALCULATIONS

### A. Structural Setup

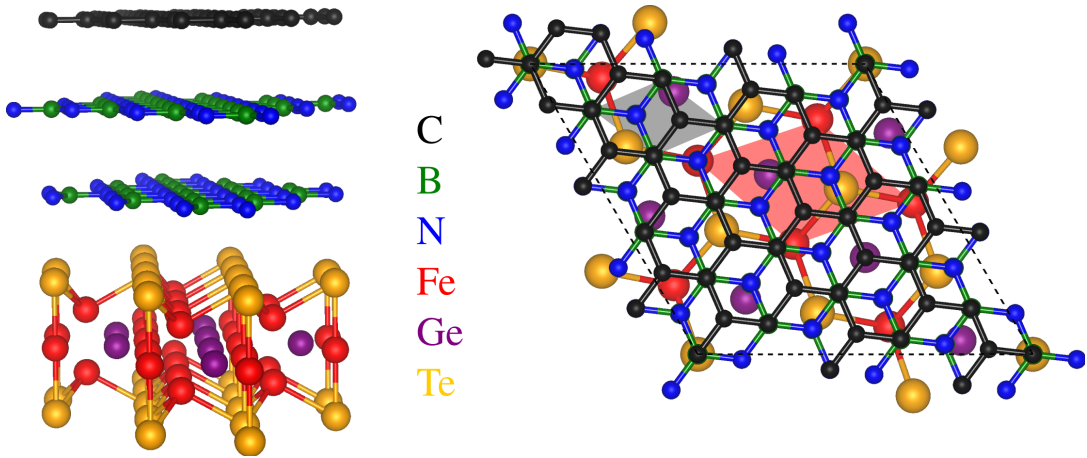

FIG. S10. Top and side view of the Fe$_3$GeTe$_2$/hBN/hBN/graphene heterostructure. The supercell has 156 atoms, with the lattice constant of 10.559 Å. The relaxed average interlayer distance between the lower hBN layer and the Te atoms of the Fe$_3$GeTe$_2$ layer is $d = 3.161$ Å. The remaining interlayer distances are fixed to 3.3 Å, and we employ an AA' stacking of the hBN layers. The grey (red) shaded area indicates the monolayer graphene/hBN (Fe$_3$GeTe$_2$) unit cell.

The Fe$_3$GeTe$_2$/hBN/hBN/graphene heterostructure was set-up with the `atomic simulation environment (ASE)` [2] and the `CellMatch` code [3], implementing the coincidence lattice method [4, 5]. The lattice constant of Fe$_3$GeTe$_2$ within the heterostructure is 3.991 Å, following the literature value [6], while the graphene and hBN layers are biaxially strained to a lattice constant of 2.423 Å. Therefore, the individual monolayers are barely strained in our heterostructure and we should be able to reliably extract band offsets as well as proximity exchange effects on Dirac states. In order to simulate quasi-2D systems, we add a vacuum of about 20 Å to avoid interactions between periodic images in our slab geometry. The resulting heterostructure is shown in Fig. S10.

### B. Computational Details

The electronic structure calculations and structural relaxations of the Fe$_3$GeTe$_2$/hBN/hBN/graphene heterostructure is performed by DFT [7] with `Quantum ESPRESSO` [8]. Self-consistent calculations are carried out with a $k$-point sampling of $24 \times 24 \times 1$. We perform open shell calculations that provide the spin-polarized ground state of the Fe$_3$GeTe$_2$ monolayer. We use an energy cut-off point for charge density of 1400 Ry and the kinetic energy cut-off for wavefunctions is 130 Ry for the scalar relativistic pseudopotentials with the projector augmented wave method [9] with the Perdew-Zunger local density approximiation (LDA). Our choice for LDA is based on Ref. [10], since calculated magnetic moments are close to experimental values. In fact, we find about 1.69 $\mu_B$/Fe, while experiments find 1.625 $\mu_B$/Fe [11] For the relaxation of the heterostructures, we add DFT-D2 vdW corrections [12–14] and use quasi-Newton algorithm based on trust radius procedure. The atoms of the Fe$_3$GeTe$_2$ layer are allowed to move freely within the heterostructure geometry during relaxation, while graphene and hBN atoms are kept fixed. Relaxation is performed until every component of each force is reduced below $1 \times 10^{-3}$ [Ry/$a_0$], where $a_0$ is the Bohr radius.

For the interpretation of the spin injection, it is helpful to know about the spin polarization of Fe$_3$GeTe$_2$. A measure for the degree of spin polarization is the tunneling density of states (TDOS), which is defined via the product of the DOS and the velocity of the Bloch bands [15]. For that purpose, we consider bulk Fe$_3$GeTe$_2$ with lattice parameters from experiment [6]. The TDOS is calculated by employing the tetrahedron method on a $27 \times 27 \times 27$ $k$-mesh, where we postprocess the `Quantum ESPRESSO` output to calculate Bloch band velocities. The remaining calculation parameters are the same as above.

### C. Results

In Fig. S11 we show the calculated band structure of the $Fe_3GeTe_2$/hBN/hBN/graphene heterostructure. Due to the two hBN layers, the Dirac states of graphene are preserved within the heterostructure. However, there is quite some charge transfer present, as the Dirac point is located about 215 meV below the heterostructure Fermi level. We do not find any signature of proximity induced exchange coupling in the graphene layer, as no magnetic moments are induced and the Dirac bands are not spin split. The opening of an orbital gap is due to the hBN layers, which introduces sublattice asymmetry and a staggered potential gap [16] Calculated spin-resolved DOS of $Fe_3GeTe_2$ and the corresponding TDOS are shown in the main text in Fig. 6.

The video file in the supplementary materials shows the spin resolved evolution of the density of states in the Brillouin Zone around the Fermi-level for bulk FGT.

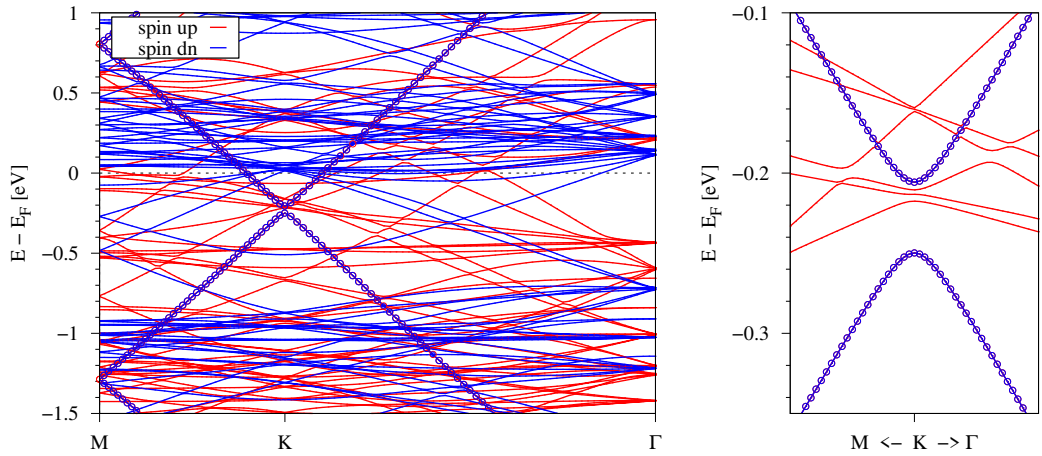

FIG. S11. Left: DFT-calculated band structure of the $Fe_3GeTe_2$/hBN/hBN/graphene heterostructure. Red (blue) lines correspond to spin up (down) and the open spheres are projections onto graphene states. Right: Zoom to the Dirac states at K. The hBN layers prevent a proximity exchange splitting of Dirac bands.

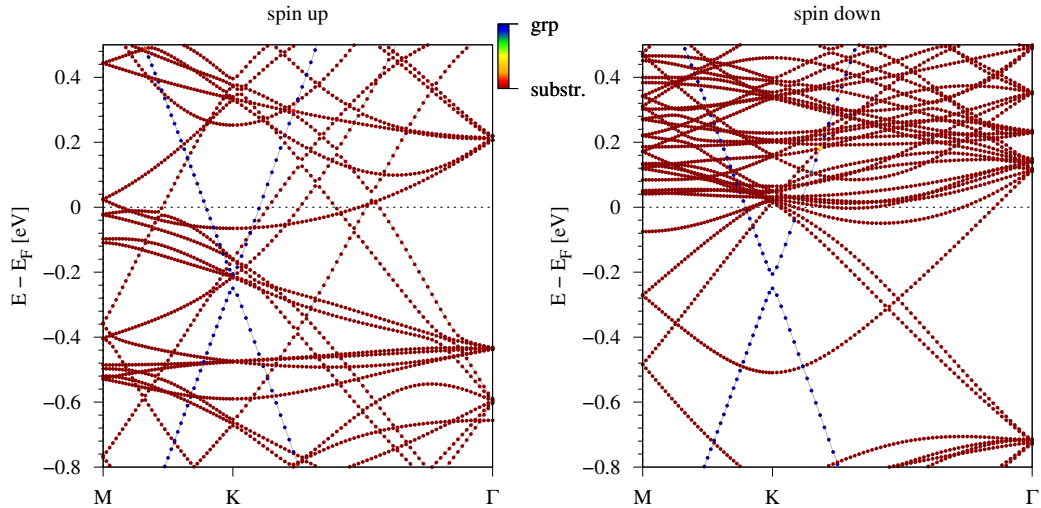

FIG. S12. DFT-calculated spin-resolved band structure of the $Fe_3GeTe_2$/hBN/hBN/graphene heterostructure. The color code represents the projection onto the graphene or substrate states.



\end{document}

